\documentclass[aps,prl,twocolumn]{revtex4}
\usepackage{epsf}
\begin{document}
\title{Decoherence and recoherence from vacuum fluctuations near a conducting plate}
\author{Francisco D.\ Mazzitelli$^{1}$, Juan Pablo Paz$^{1,2}$,  
Alejandro Villanueva$^1$}

\affiliation{(1): Departamento de F\'\i sica {\it J.J. Giambiagi}, FCEyN UBA, 
Ciudad Universitaria, Pabell\' on I, 1428 Buenos Aires, Argentina}
\affiliation{(2): Theoretical Division, MS B213 LANL, Los Alamos,
NM 87545, USA}
\date{\today}

\begin{abstract}

{The interaction between particles and the electromagnetic field induces
decoherence generating a small suppression of fringes in an
interference experiment. We show that if a double slit--like
experiment is performed in the vicinity of a conducting plane, the fringe
visibility depends on the position (and orientation) of the
experiment relative to the conductor's plane. This phenomenon is
due to the change in the structure of vacuum induced
by the conductor and is closely related to the Casimir effect.
We estimate the fringe visibility both for charged and for neutral
particles with a permanent dipole moment. The 
presence of the conductor may tend to increase decoherence in
some cases and to reduce it in others. A simple explanation for
this peculiar behavior is presented.}
\end{abstract}


\maketitle

The interaction of a quantum system with its environment is responsible
for the process of decoherence, which is one of the main ingredients to 
understand the quantum--classical transition \cite{decoherence}. In some
cases, the interaction with the environment cannot be 
switched off. This is the case for charged particles that unavoidably
interact with the electromagnetic field. As this interaction
is fundamental, its effect is present in any interference experiment.
In this letter we will analyze the influence of a conducting boundary 
in the decay of the visibility of interference fringes in a double slit 
experiment performed with charged particles (or neutral particles with a 
dipole moment). The reduction of fringe visibility is induced by the 
interaction between the particles and the electromagnetic field. 
Some aspects of this problem have been analyzed before. In fact, 
it is known that for charged particles, the interaction between the system
(the particle) and the environment (the electromagnetic field)
induces a rather small decoherence effect 
even if the initial state of the field is the vacuum
\cite{stern1,stern2,barone,breuer,vourdas,ford,hu}. 
A particularly simple expression for the decay in the fringe visibility 
was obtained in \cite{stern1,stern2}:
Assuming an electron in harmonic motion (with
frequency $\Omega$) along the relevant trajectories of the double slit
experiment, the fringe visibility decays by a factor $(1-P)^2$
where $P$ is the probability that a dipole $p=eR$ oscillating at
frequency $\Omega$ emits a  photon ($R$ is the characteristic size of the trajectory).
This result is in accordance with the idea that decoherence becomes effective
when a record of the state of the system is irreversibly imprinted in
the environment. In this case, after photon emission, if the electron
follows the trajectory $\vec X_1(t)$ of the double slit
experiment (see Figure 1) it becomes correlated with a state of the
environment $|E_1(t)\rangle$. This state is different from the one with
which the electron correlates if it follows the trajectory $\vec X_2(t)$. 
The absolute value of the overlap between these two different states 
is precisely given by $(1-P)^2$.

In this letter we will analyze how the fringe visibility is modified when  
performing a double slit interference experiment in the vicinity of a
conducting plane. Our analysis will serve not only to correct 
previous results \cite{ford} but also to show that the 
effect of the conductor is quite remarkable and simple to understand. 
As we will see, the presence of the conducting plane may produce more
decoherence in some cases and less decoherence in others. For example, we
will show that if a conducting plate is placed perpendicular to the
trajectories of the interfering charge, the fringe visibility decreases with
respect to the vacuum case (absence of conducting plate). However, if the
plate lies parallel to the electron's trajectories, the contrast increases
(the system recoheres!). We will show that this peculiar behavior can be
understood in simple terms and the magnitude of the effect can be easily 
estimated.
There are several interesting physical effects connected with the one we 
are analyzing here. Thus, it is well known that a conducting boundary modifies 
the properties of the zero point fluctuations, and therefore could affect 
the interference experiments of particles that interact with the 
electromagnetic field. Other consequences of the presence of 
nontrivial boundary conditions are the Casimir force between
two conductors \cite{casimir} and the Casimir--Polder force \cite{casimir-polder}
affecting a probe particle in the vicinity of a conductor.
These phenomena, that have been experimentally verified
\cite{casimir-experiments}, are close relatives of the process we are studying
here. In fact, the Casimir--Polder force can be thought as the dispersive
counterpart of the decoherence effect we will discuss. The influence of boundaries
on the electromagnetic vacuum is also responsible for changes in atomic lifetimes
and interference phenomena for light emitted by atoms near conducting surfaces
\cite{atomic-experiments}.

\begin{figure}
 \centering \leavevmode
 \epsfxsize 3.2in
 \epsfbox{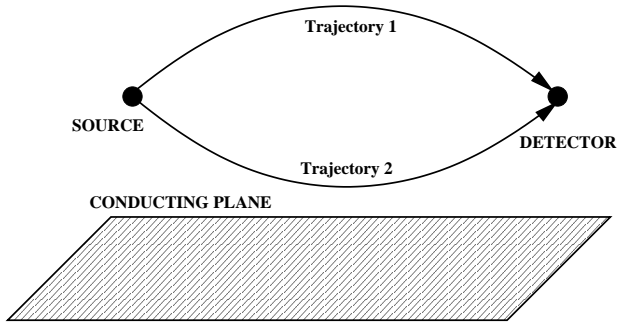}
\vspace{0.25 cm}
\caption{Scheme for a double slit like experiment near a conducting plane. 
The component of the velocity in the direction from the source to the detector
is assumed to be constant.}
\label{fig1}
\end{figure}

Let us first outline a simple method to compute the effect of electromagnetic 
interactions on the fringe contrast. 
We consider two electron wave packets that follow well defined
trajectories $\vec{X} _1(t)$ and $\vec{X} _2(t)$ that coincide at initial ($t=0$)
and final ($t=T$) times as shown in Figure 1. In the absence of environment, the interference
depends on the relative phase between both wave packets at $t=T$. Because of the
interaction with the quantum electromagnetic field, the interference pattern is affected. This
effect can be calculated as follows: We  assume an initial state of the combined particle--field
system of the form $|\Psi(0)\rangle= (|\phi_1\rangle+|\phi_2\rangle)\otimes|E_0\rangle$. Here
$|E_0\rangle$ is the initial (vacuum) state of the field and
$|\phi_{1,2}\rangle$ are two states of the electron that are localized around the initial point
and that in the absence of other interaction continue to be localized along the trajectories
$X_{1,2}(t)$ respectively. At later times, due to the particle field interaction the state of
the combined system becomes
$|\Psi(t)\rangle= (|\phi_1(t)\rangle\otimes |E_1(t)\rangle+|\phi_2(t)\rangle
\otimes|E_2(t)\rangle)$.
Thus, the two localized states $|\phi_1(t)\rangle$
and $|\phi_2(t)\rangle$ become correlated with two different states of the field. Therefore, the
probability of finding a particle at a given position turns out to be
\begin{eqnarray}
prob(\vec X,t)&=&|\phi_1(\vec X,t)|^2+|\phi_2(\vec X,t)|^2 \nonumber\\
&+& 2{\rm Re}(\phi_1(\vec X,t)\phi_2^*(\vec X,t)\langle E_2(t)|E_1(t)\rangle).
\label{probability}
\end{eqnarray}

The overlap factor $F=\langle E_2(t)|E_1(t)\rangle$ is responsible for two effects: Its phase
produces a shift of the interference fringes. 
The absolute value $|F|$ is responsible for the decay in the fringe contrast, 
which is the phenomenon we will analyze here. The calculation of
the factor $F$ is conceptually simple since it is nothing but the overlap between
two states of the field that arise from the vacuum under
the influence of two different sources (this factor is identical to the Feynman--Vernon
influence functional \cite{Feynman-Vernon}). Each of the two states of the field
can be written as
$|E_a(t)\rangle=T\left(\exp(-i\int d^x J_a^\mu (x)A_\mu (x))\right)|E_0\rangle$,
where $J_a^\mu (x)$ is the conserved
$4$--current generated by the particle following the classical trajectory $\vec X_a(t)$,
i.e. $J_a^\mu (\vec X,t)=\left(e,e\dot{\vec X}_a (t)\right)\times\delta^3(\vec X-\vec X_a(t))$,
($a=1,2$). Using this, it is simple to derive an expression for the absolute value of the
overlap. Denoting $|F|=\exp(-W_c)$, we get 
\begin{equation}
W_c={1\over 2}\int d^4x\int d^4y (J_1-J_2)^\mu(x)D_{\mu\nu}(x,y)(J_1-J_2)^\nu(y),
\label{Wcharges}
\end{equation}
where $D_{\mu\nu}$ is the expectation value of the
anti--commutator of two field operators: $D_{\mu\nu}(x,y)={1\over
2}\langle\{A_\mu(x),A_\nu(y)\}\rangle$. It is easy to show that
the square of the overlap has a simple interpretation: $|F|^2$ is
equal to the probability for vacuum persistence in the presence of
a source $j_{\mu}=(J_1-J_2)_\mu$, which corresponds to
a time dependent electric dipole ${e}(\vec X_1(t)-\vec X_2(t))$ \cite{breuer}.

A conceptually
similar and physically interesting problem can be analyzed along the same lines: the
decoherence of neutral particles with a non--vanishing permanent dipole moment. In such
case we can model the particle--field interaction 
using a Lagrangian $L_{int}=P_{\mu\nu}(x)F^{\mu\nu}(x)$. Here $F_{\mu\nu}$ is the field
strength tensor and $P_{\mu\nu}$ is a totally antisymmetric tensor whose non--vanishing
components are given in terms of the electric and magnetic dipole densities. For particles
with electric dipole $\vec p$ and magnetic dipole $\vec m$ moving along a trajectory
$\vec X(t)$, the dipolar tensor is such that $P_{0i}=p_i(t)\delta^3(\vec X-\vec X(t))/2$
and $P_{ij}=\epsilon_{ijk}m_k (t)\delta^3(\vec X-\vec X(t))/2$.
In this case we can perform a calculation which is similar to the one above and
show that the overlap $F=\exp(-W_d)$ is 
\begin{eqnarray}
W_d&=&{1\over 2}\int\int d^4x d^4y \ (P_1-P_2)^{\mu\nu}(x)K_{\mu\nu\rho\sigma}(x,y)\times\nonumber\\
&\times& (P_1-P_2)^{\rho\sigma}(y),
\label{Wdipole}
\end{eqnarray}
where the kernel is 
$K_{\mu\nu\rho\sigma}(x,y)=\langle\{F_{\mu\nu}(x),F_{\rho\sigma}(y)\}\rangle$.

In what follows we will present results for the {\sl decoherence factors}
$W_c$ and $W_d$ (the subscripts stand for ''charges'' and ''dipoles'').
To compute $W_c$ we need the two point function appearing in (\ref{Wcharges}).
In the Feynman gauge and in the absence of conducting plates it is
\begin{equation}
D^{(0)}_{\mu\nu}(x,y)=-\eta_{\mu\nu}\int {d^3\vec k\over{(2\pi)^32k}}
\ {\rm e}^{i\vec k(\vec x-\vec y)}\cos(k(x_0-y_0)),
\label{Dmunu}
\end{equation}
where the superscript $(0)$ identifies this as the vacuum contribution. We will assume
that the trajectories are such that $\vec X_1(t)=-\vec X_2(t)=x(t)
\hat{\rm x}$. This is
enough to describe a typical double slit experiment from the point of
view of an observer moving at constant velocity from the source to the detector.
In such case we  obtain the following relatively simple expression for $W_c$:
\begin{equation}
W_c^{(0)}=e^2 \int {d^3\vec k\over 8\pi^3 k}(1-{k_j^2\over k^2})\vert
\int_{-\infty}^{\infty} dt\,\dot x(t) \,\cos[k_x x(t)] \, e^{i k t}\vert^2.
\label{Wc0}
\end{equation}
This result, obtained first in \cite{stern1}, can be simplified
further by assuming the validity of the dipole approximation
(which is consistent in the nonrelativistic limit). 
Doing this, one can evaluate the decoherence factor for 
some special trajectories. In fact, for adiabatic trajectories,
where $x(t)= R \exp [-t^2/T^2]$, we find that $W_c^{(0)}=2e^2
v^2/3\pi$, where $v=R/T$ is a characteristic velocity. This result
is finite and free of any cutoff dependence. However, for 
trajectories evolving over a finite time the situation is
different. Thus, assuming that the motion starts at $t=0$, ends at
$t=T$, and that is composed of periods of constant velocity $v$,
or constant acceleration $v/\tau$, we obtain a result that
diverges logarithmically when $\tau\rightarrow 0$: $W_c^{(0)}={2
e^2 v^2} \log[T/\tau]/\pi^2$ (if $\tau/T\ll 1$). 
Previous results \cite{ford,breuer} were
obtained for trajectories with discontinuous velocity using a
natural UV cutoff arising from the finite size of the electron.
The results of \cite{breuer} agree with ours if the high frequency
cutoff is identified with $1/\tau$.
Thus, the cutoff
dependence disappears in the adiabatic case and is a consequence
of abrupt changes in velocity and the instantaneous preparation of
the initial state. If the two wave packets are superposed after
oscillating $N$ times, it is possible to define a 
decoherence rate (the ammount by
which the decoherence factor grows in a single oscillation). Thus, if the time
to complete one oscillation is much shorter than the period between oscillations
we can show that, for large $N$, the decoherence factor is
proportional to $N$: $W_c^{(0)}=N W_c^{(0)}(1)$
where $W_c^{(0)}(1)$ is the decoherence factor in a single oscillation.

We will now show how this result is modified by the presence of a perfect
conductor located in the plane $z=0$. To consider the effect of the conductor
we only need to use the appropiate two point function $D_{\mu\nu}$ that is
the sum of two terms \cite{Brown}:
$D_{\mu\nu}=D_{\mu\nu}^{(0)}+D_{\mu\nu}^{(B)}$. The vacuum term is the same as
in (\ref{Dmunu}).
The contribution of the boundary conditions (identified by the superscript (B))
can be obtained by the method of images and is:
\begin{eqnarray}
D^{(B)}_{\mu\nu}(x,y)&=&(\eta_{\mu\nu}+2n_\mu n_\nu)\int {d^3\vec k\over{(2\pi)^32k}}
\times\nonumber\\
&\times&\exp(i\vec k(\vec x-{\vec y}'))\cos(k(x_0-y_0)).
\label{Dplate}
\end{eqnarray}
Here $n^\mu$ is the normal to the plane and ${\vec y}'$ is the position of
the image point of $\vec y$ (a prime denotes a vector reflected with respect
to the plane, i.e. ${\vec y}'=(y_x,y_y,-y_z)$). Using (\ref{Dplate}) we can derive
a formula for the contribution of the boundary to the decay of the interference fringes.
The complete equation
is involved and will be given elsewhere \cite{us-next}. Here we will restrict to the case where
the trajectories are either perpendicular or parallel to the conductor's plane.
Thus,
we will write $\vec X_{1,2}=z_0 \hat z \pm x(t) \hat {\rm \j} $ where $\hat {\rm \j} $ defines a
fixed vector aligned either along the $\hat z$--axis or along the plane perpendicular to it.
In such case, the conductor's contribution to decoherence is
\begin{eqnarray}
W_c^{(B)}&=& -\,\hat {\rm \j} {\hat {\rm \j} }'\, e^2 \int {d^3\vec k\over 8\pi^3k}\,
(1-{k_j^2\over k^2})\, e^{2ik_z z_0}\times\nonumber\\
&\times& \vert\int_0^tdt'
\, \dot x(t') \, \cos[k_j x(t')] \, e^{i k t'}\vert^2.
\label{WcB}
\end{eqnarray}
The sign of $W_c^{(B)}$ is determined by the orientation of
${\hat{\rm \j}}'$ relative to $\hat{\rm \j} $. $W_c^{(0)}$ is negative when the trajectories are
parallel to the conductor's plane (since in that case ${\hat  {\rm \j}}'=\hat{\rm \j}$).
On the other
hand, $W_c^{(0)}$ is positive when the trajectories are perpendicular to the plane (where
${\hat {\rm \j}}'=-\hat{\rm \j}$). At small distances to the plane ($z_0\simeq 0$)
we can see from (\ref{WcB}) that $\vert W_c^{(B)}\vert\simeq W_c^{(0)}$.
Therefore, if the trajectories are perpendicular to the plane,
in the limit of small distances the decoherence 
factor is $W_c=W_c^{(0)}+W_c^{(B)}\simeq 2 W_c^{(0)}$: The effect of
the conductor is to double the decoherence factor.
However, if the trajectories are parallel to the conductor the effect is exactly the
opposite: As $W_c^{(B)}$ is negative, the conductor produces {\sl recoherence} increasing
the contrast of the fringes. In fact, for small distances the decoherence
factor tends to vanish since $W_c=W_c^{(0)}+W_c^{(B)}\simeq 0$.
These results can be understood using the method of images taking into account that
decoherence in empty space is related to the probability of photon emision for a
varying dipole $p=e x(t)$. When the conducting plane is parallel to the dipole, the image
dipole is $\vec p_{im}=-\vec p$. Therefore the total dipole moment vanishes, and so does
the probability to emit a photon. The image dipole cancels the effect of the real dipole
and this produces the recovery of the fringe contrast. On the other hand, when the
conductor is perpendicular to the trajectories, the image dipole is equal to
the real dipole $\vec p_{im}=+\vec p$.
Therefore, the total dipole is twice the original one. This in principle would lead
us to conclude that the total decoherence factor $W_c=W_c^{(0)}+W_c^{(B)}$ should be
four times larger than $W_c^{(0)}$. However, one should take into account that in the
presence of a perfect mirror photons can only be emitted in the $z\geq 0$ region. This
introduces an additional factor of $1/2$ that gives rise to the final result
$W_c\simeq 2 W_c^{(0)}$.
The impact of conducting boundaries on the fringe visibility for interference experiments
performed with charged particles
was previously examined in \cite{ford}. However, results obtained
in such papers are not correct due to inconsistent approximations that violate
the conservation of the $4$--current. Thus the expressions obtained there differ
from ours in several ways: not only they are not proportional to $v^2$ but also they
violate the possitivity of the total decoherence factor $W_c$.

Let us now describe the results for the case of neutral particles with permanent dipole
moments. The calculation is tedious and details 
will be given elsewhere \cite{us-next}.
Here we will analyze it under somewhat simplified assumptions. Will assume that
the dipole moments $\vec p$ and $\vec m$ remain constant along the trajectories. If
$\vec X_1(t)=-\vec X_2(t)=x(t)\hat{\rm \j}$ we find:
\begin{eqnarray}
W_d^{(0)}&=& \int {d^3\vec k\over 8\pi^3}\,k
\{\vec p^2(1-{k_p^2\over k^2})+\vec m^2(1-{k_m^2\over k^2})\times\nonumber\\
&\times&\vert\int_0^tdt'\sin[k_j x(t')] \, e^{i k t'}\vert^2.
\label{Wd0}
\end{eqnarray}
Using again the dipole approximation, for the adiabatic trajectory
the decoherence factor is such that
$W_d^{(0)}/W_c^{(0)}\simeq p^2/e^2T^2$ for a purely electric dipole. This ratio
is typically much smaller than one.

The effect of the conductor can also be taken into account using the method
described above. For simplicity we will only consider trajectories that are parallel
to the plane (i.e., $\vec X_{1,2}=z_0\hat z \pm x(t) \hat {\rm \j}$) and assume that
the dipole moments are either perpendicular or parallel
to the conductor (the general case is more complex but the essential features
can be seen here). Using this we obtain
\begin{eqnarray}
W_d^{(B)}&=& -\, \int {d^3\vec k\over 32\pi^3}\,k
\{\vec p {\vec p}'(1-{k_p^2\over k^2})-\vec m {\vec m}'(1-{k_m^2\over k^2})\,
\times\nonumber\\
&\times& e^{2ik_zz_0}\vert\int_0^tdt'\sin[k_j x(t')] \, e^{i k t'}\vert^2.
\label{WdB}
\end{eqnarray}

Thus, if the reflected dipole ${\vec p}'$ has the opposite 
direction than $\vec p$
(which is the case when $\vec p$ is parallel to the plate) the conductor tends
to increase decoherence (since the contribution of the electric 
dipole to $W_d^{(B)}$
is positive). Likewise, when $\vec p$ is perpendicular
to the plane, $\vec p={\vec p}'$ and
the contribution of the electric dipole to $W_d^{(B)}$ is negative. Therefore, in this
case the conductor produces {\sl recoherence} instead of decoherence.
The opposite effect is found for the magnetic dipole. Indeed, when
$\vec m'=-\vec m$ (magnetic dipole perpendicular to the plane) the conductor produces
recoherence while more decoherence is produced if the magnetic dipole is parallel to the
plane. This features can also be understood by thinking in terms of the image dipoles
that are generated by the conductor. Thus, both when the $\vec p$ is perpendicular
to the plane or when $\vec m$ is parallel, the direction of the image dipoles
coincide with the source dipoles. In such case the decoherence increases. In
the opposite situation ($\vec p$ parallel or $\vec m$ perpendicular to
the plane) the effect of the conductor is to introduce recoherence.
Again, in the limit of small distances the absolute value of $W_d^{(0)}$ and $W_d^{(B)}$
coincide and therefore the decoherence factor doubles with respect to the vacuum case.

For the two cases we considered (charges and dipoles) one can show that the boundary
contribution to the decoherence factor decays algebraically with the distance to the
conductor (in the limit of large distances). For small separations, explicit expressions
for the decoherence factor can be obtained. For example, for charges moving close
and parallel to the conductor, the lowest order contribution of
$W_c$ depends quadratically on $z_0$. As expected, it
exactly coincides with the decoherence factor produced by an electric dipole
$p=2ez_0$ in vacuum (with an additional factor of $1/2$ that takes into account
that photons can only be emitted with $z\geq 0$).

In conclusion, our work
shows that the effect of conducting boundaries on interference experiments
has a rather simple interpretation: The way in which decoherence is affected
is similar to the manner in which atomic emission properties are modified by the presence of
conducting boundaries. Thus, the effect of the boundaries has not a well defined
sign and may produce either more decoherence or complete recoherence (i.e. smaller or
higher fringe visibility than in vacuum) depending on the orientation of the relevant
trajectories with respect to the conductor's plane. The effect discussed here
is conceptually important due to its fundamental origin (i.e., it is always present)
but its magnitude is too small to be under the reach of current experiments involving
interference of neutral atoms in the vicinity of conducting planes \cite{atom-chip}.

This work  was supported by UBA, CONICET, Fundaci\'on Antorchas
and ANPCyT, Argentina.
F.D.M thanks ICTP for hospitality during completion of this work.

\end{document}